\documentclass[aps,prl,preprint,groupedaddress]{revtex4}

\usepackage{graphicx}

\def\barP{\overline P}
\def\barT{\overline T}
\def\barH{\overline H}
\def\barPsi{\overline\Psi}
\def\barPi{\overline\Pi}

\def\PRD#1#2#3{Phys. Rev. {\bf D#1}, #2 (#3)}
\def\NPB#1#2#3{Nucl. Phys. {\bf B#1}, #2 (#3)}

\def\PRL#1#2#3{Phys. Rev. Lett. {\bf #1}, #2 (#3)}
\def\PRep#1#2#3{Phys. Rep. {\bf #1}, #2 (#3)}

\begin{document}

\preprint{
OCHA-PP-301
}

\title{
Realistic model for SU(5) grand unification
}


\author{
Noriyuki Oshimo
}
\affiliation{
Graduate School of Humanities and Sciences {\rm and} Department of Physics \\
Ochanomizu University, Tokyo, 112-8610, Japan
}


\date{\today}

\begin{abstract}
     A grand unified model based on SU(5) and supersymmetry is presented.  
Pairs of superfields belonging to $\bf 15$ and $\overline{\bf 15}$ representations 
are newly introduced, while gauge coupling unification is satisfied.  
Improper mass relations in the minimal model between charged 
leptons and $d$-type quarks are corrected.  
Neutrinos have non-vanishing masses, with large angles for generation mixings 
of the leptons being compatible with the small angles of the quarks.  
Constraints from proton decay are relaxed.  
A new source for lepton number generation in the early universe is provided.  

\end{abstract}


\maketitle


     Masses and generation mixings of quarks and leptons provide useful 
information on the models which assume grand unification.  
In grand unified theories (GUTs), some quarks and some leptons are 
contained in one multiplet of the gauge group, coupling to 
a Higgs boson with the same magnitude.   
Masses and generation mixings for these quarks and leptons 
are thus correlated with each other.  
Given the correlations at the GUT energy scale, those 
at the electroweak energy scale are predicted.  
These predictions are confronted with experimental 
measurements, giving nontrivial constraints on the models.   

     In the models whose gauge group is SU(5) or SO(10), a serious  
problem has been known for a long time on the masses of 
$d$-type quarks and charged leptons \cite{langacker}.  
If particle contents are minimal, their coefficient matrices for Higgs couplings  
have the same eigenvalue in each generation at the GUT energy scale.  
However, the observed masses are not consistent with this GUT relation.   
Another problem has also been recognized these years by neutrino 
experiments for masses and generation mixings \cite{chen}.  
In SU(5) models, right-handed neutrinos belong to singlet group representation, 
so that their existence for non-vanishing masses would be considered {\it ad hoc}.  
In SO(10) models, generation mixings for quarks and leptons described by the 
Cabbibo-Kobayashi-Maskawa (CKM) matrix and 
the Maki-Nakagawa-Sakata (MNS) matrix are correlated.    
Measured large mixing angles of the MNS matrix cannot coexist 
trivially with the small mixing angles of the CKM matrix.  

     In this letter we present an SU(5) model which can describe nature realistically.  
Particle contents are enlarged, though their   
group representations are within rank two, the same as the minimal model.  
Supersymmetry is imposed in order 
to embed consistently the standard model in the framework of GUTs.  
In the SU(5) model, even if right-handed neutrinos are not included, small 
neutrino masses could be generated by introducing Higgs superfields 
of representations $\bf 15$ and $\overline{\bf 15}$  \cite{oshimo}.  
We argue that masses and mixings of quarks and leptons can all be 
described by introducing superfields of these representations.  
Incorporation of the new particles does not ruin gauge coupling unification of 
SU(3), SU(2), and U(1) which is achieved in the minimal supersymmetric SU(5) model.  
In addition, constraints on the model from proton decay, which 
are severe in the minimal model, are loosened.  
Non-vanishing lepton number is induced by decays of new particles 
in the early universe, which could be the origin of present baryon asymmetry.   

     The model consists of Higgs superfields $\Phi$, $H$, $\barH$, $T_m$, $\barT_m$ 
($m$=1,2) and matter superfields $\Psi_i$, $\barPsi_i$ ($i$=1,2,3), $\Pi$, $\barPi$.  
Their representations for SU(5) are shown in Table \ref{particles}.   
We assume that each superfield has an intrinsic parity: even for Higgs and odd for matter.  
Newly introduced particles are pairs of $\bf 15$ and $\overline{\bf 15}$, 
{\it i.e.}, two pairs for Higgs and one pair for matter.  
The model is free from gauge anomaly.    
Under SU(3)$\times$SU(2)$\times$U(1) symmetry, $\bf 15$ representation is 
decomposed into $({\bf 6},{\bf 1}, -2/3)$, $({\bf 3},{\bf 2},1/6)$, and $({\bf 1},{\bf 3},1)$, 
where U(1) charges are expressed as hypercharges.  

     The superpotential prescribed by SU(5) and intrinsic parity is given by 
\begin{eqnarray}
    W &=& W_H + W_{M1} + W_{M2} + W_{M3},  
     \nonumber \\
    W_H &=& M_H{\barH}H + M_{Tm}{\barT_m}T_m
          + \frac{1}{2}M_\Phi\Phi^2 
                            \nonumber \\
      &+&  \lambda_{H\Phi} {\barH}\Phi H 
          + \lambda_{T\Phi}^{mn} \barT_m\Phi T_n 
          + \frac{1}{3}\lambda_\Phi\Phi^3 
                            \nonumber  \\
      &+&  {\overline{\lambda}_{HT}^{m}}{\barH}T_m{\barH}
          + \lambda_{HT}^{m}H{\barT_m}H, 
\label{superpotential} \\
    W_{M1} &=& M_\Pi \barPi\Pi 
          + \chi_\Pi \barPi\Phi\Pi  
          + \chi^{i}_{\Pi\Psi} \barPi\Phi\Psi_i,  
  \nonumber  \\
    W_{M2} &=& \eta^{i}_{\Pi\Psi}{\barH}\Pi{\barPsi_i}    
      + \eta^{ij}_{\Psi}\epsilon_5 H\Psi_i\Psi_j  
         + {\overline\eta}^{ij}_{\Psi}{\barH}\Psi_i{\barPsi_j},  
  \nonumber  \\
    W_{M3} &=& \kappa^{mij}_{\Psi} {\barPsi_i}T_m{\barPsi_j},  
  \nonumber  
\end{eqnarray}
where contraction of SU(5) group indices is understood.  
We write the totally antisymmetric tensor of rank $n$ as $\epsilon_n$.  
The mass terms of $T_m$ and $\barT_m$ can be taken for 
diagonal without loss of generality.  
The mass parameters $M_H$, $M_{Tm}$, $M_\Phi$, and $M_\Pi$ have magnitudes 
around the order of the GUT energy scale $M_X$, which is given 
typically by $M_X\sim 10^{16}$~GeV.  
The superpotential $W_H$ contains only Higgs superfields.  
The couplings of matter and Higgs superfields are contained in the superpotentials 
$W_{M1}$, $W_{M2}$, and $W_{M3}$ .  
The coefficients $\eta^{ij}_{\Psi}$ and $\kappa^{mij}_{\Psi}$ are 
symmetric for the indices $i$ and $j$.  

\begin{table}
\caption{
Superfields and their representations for SU(5), $m$=1,2 and $i$=1,2,3.  
\label{particles}
}
\begin{ruledtabular}
\begin{tabular}{ccccc}
$\Phi$   & $T_m$   & $\barT_m$   & $H$   & $\barH$ \\
$\bf 24$ & $\bf 15$ & $\overline{\bf 15}$ & $\bf 5$ & $\overline{\bf 5}$ \\
\hline
$\Pi$ & $\barPi$ & $\Psi_i$ & $\barPsi_i$ &  \\ 
$\bf 15$ & $\overline{\bf 15}$ & $\bf 10$ & $\overline{\bf 5}$ &  \\
\end{tabular}
\end{ruledtabular}
\end{table}

     The vacuum of this model is determined by the superpotential $W_H$ 
in Eq. (\ref{superpotential}).  
The SU(5) gauge symmetry is broken spontaneously by 
$\Phi$ of adjoint representation.  
Its scalar component $\tilde\Phi$ has a vacuum expectation value (VEV) 
$\langle \tilde\Phi \rangle= {\rm diag}(1, 1, 1, -3/2,-3/2)v_\Phi$,  where $v_\Phi$ is 
given by $v_\Phi\simeq 2{\rm Re}(M_\Phi)/{\rm Re}(\lambda_\Phi)$. 
The gauge bosons $X$ and $Y$ then have masses, 
$M_X^2=M_Y^2=(25/8)g_5^2v_\Phi^2$, with $g_5$ being the 
SU(5) gauge coupling constant.  
The residual gauge symmetry is  
SU(3)$\times$SU(2)$\times$U(1) of the standard model.   

     Below the GUT energy scale, the SU(3)-singlet components of $H$, $\barH$, $T_m$, 
$\barT_m$, and $\Phi$ are responsible for further breaking of gauge symmetry.   
The relevant part of the superpotential $W_H$ is written as 
\begin{eqnarray}
    W &=& -m_H H_1\epsilon_2 H_2 + m_{Tm}\barT_m T_m  
          + \frac{1}{2}m_\Phi \Phi^2 
                            \nonumber \\
      &+&  \lambda_{\phi 1} (\epsilon_2 H_1)\Phi H_2 
          + \lambda_{\phi 2}^{mn} \barT_m\Phi T_n 
          + \frac{1}{3}\lambda_{\phi 3}\Phi^3 
                            \nonumber \\
    &+& \overline{\lambda}^m(\epsilon_2 H_1) T_m\epsilon_2 H_1 
          + \lambda^m H_2\barT_mH_2. 
\label{superpotentialH}  
\end{eqnarray}
Here, $H_1$ and $H_2$ stand for components of $\barH$ and $H$, 
which belong respectively to $({\bf 1},{\bf 2},-1/2)$ and $({\bf 1},{\bf 2},1/2)$. 
The SU(2)-triplet components of $T_m$, $\barT_m$, and $\Phi$ are denoted 
by the same symbols, though $T_m$ and $\barT_m$ express linear 
combinations of original superfields.
At SU(5) breaking, the mass parameters $m_{Tm}$ are given by 
${\rm diag}(m_{T1}, m_{T2})=\bar A^\dagger[{\rm diag}(M_{T1},M_{T2})-(3/2)\lambda_{T\Phi}v_\Phi]A$, 
where $\bar A$ and $A$ represent $2\times 2$ unitary matrices.     
The other mass parameters are given by  
$m_H=M_H-(3/2)\lambda_{H\Phi}v_\Phi$ and $m_\Phi=M_\Phi-3\lambda_\Phi v_\Phi$.  
The coefficients are given by $\lambda_{\phi 1}=\lambda_{H\Phi}$, 
$\lambda_{\phi 2}=\bar A^\dagger\lambda_{T\Phi}A$, 
$\lambda_{\phi 3}=\lambda_\Phi$, $\overline{\lambda}^m=A_{lm}{\overline\lambda}_{HT}^l$, 
and $\lambda^m=\bar A^*_{lm}\lambda_{HT}^l$.  
We have renamed $A^*_{lm}T_l$ and $\bar A_{lm}\barT_l$ as $T_m$ and 
$\barT_m$, respectively.  
The magnitude of $m_H$ is assumed of the order of 
the electroweak energy scale $M_W$.  
The mass parameters $m_{Tm}$ and $m_\Phi$ have magnitudes around 
the order of $M_X$.  

     The SU(2)$\times$U(1) gauge symmetry is broken spontaneously 
down to U(1)$_{\rm EM}$ by the SU(2)-doublet superfields $H_1$ and $H_2$.  
Non-vanishing VEVs are induced for their scalar components, 
$\langle \tilde H_1\rangle =(v_1/\sqrt{2}, 0)$ and $\langle \tilde H_2\rangle =(0, v_2/\sqrt{2})$, 
where $v_1$ and $v_2$ are related to the $W$-boson mass as 
$M_W^2=(1/4)g_2^2(v_1^2+v_2^2)$, with $g_2$ being the SU(2) gauge coupling constant.  
 
     The symmetry breaking of SU(2)$\times$U(1) induces non-vanishing VEVs for 
the scalar components of $T_m$ and $\barT_m$, as well as $\Phi$.  
Since these superfields have large masses, magnitudes of the VEVs are 
extremely smaller than $v_1$ and $v_2$.  
Assuming that U(1)$_{\rm EM}$ symmetry is not broken, 
the VEVs are obtained as $\langle \tilde T_m \rangle={\rm diag}(0,v_{Tm}/\sqrt{2})$ 
and $\langle \tilde{\barT}_m \rangle = {\rm diag}(0,v_{\barT m}/\sqrt{2})$, with 
\begin{eqnarray}
 |v_{Tm}| = \left|\frac{\lambda^m v_2^2}{\sqrt{2}m_{Tm}}\right|, 
\quad 
 |v_{\barT m}| = \left|\frac{\overline{\lambda}^mv_1^2}{\sqrt{2}m_{Tm}}\right|.  
\label{vevT}
\end{eqnarray}
For $\lambda^m, \overline{\lambda}^m \sim 1$ and $m_{Tm}\sim 10^{14}$~GeV, 
the values of $v_{Tm}$ and $v_{\barT m}$ are of the order of $10^{-1}$~eV, 
which is the same order of magnitude as observed neutrino masses.  

     Now we discuss masses and mixings of matter superfields.  
Under SU(5) symmetry, the masses of $\Pi$ and $\barPi$ are given by $M_\Pi$, 
while those of $\Psi_i$ and $\barPsi_i$ are vanishing.  
However, both $\Psi_i$ and $\Pi$ have a component of the same representation 
$({\bf 3},{\bf 2},1/6)$, which are denoted respectively by $P_i$ ($i$=1,2,3) and $P_4$.  
At SU(5) breaking, these matter superfields are mixed.   
The superpotential $W_{M1}$ in Eq. (\ref{superpotential}) gives mass terms, 
\begin{eqnarray}
W &=& M_{Q4} \barP_4F_{4j} P_j+M_{(6)}\barPi_{(6)}\Pi_{(6)}+M_{(1)}\barPi_{(1)}\Pi_{(1)},      
    \nonumber \\
M_{Q4}^2 &=& \sum_{i=1}^3\left|\frac{5}{4}\chi^{i}_{\Pi\Psi} v_\Phi\right|^2 
              + \left| M_\Pi-\frac{1}{4}\chi_\Pi v_\Phi\right|^2,    
                \\
  F_{4i} &=& -\chi^i_{\Pi\Psi} \frac{5v_\Phi}{4M_{Q4}},  
            \quad 
  F_{44} = \frac{M_\Pi}{M_{Q4}}-\chi_\Pi \frac{v_\Phi}{4M_{Q4}},    
     \nonumber \\
    M_{(6)}&=&M_\Pi+\chi_\Pi v_\Phi,     \quad  M_{(1)}=M_\Pi-\frac{3}{2}\chi_\Pi v_\Phi,  
         \nonumber 
\end{eqnarray}
where $\barP_4$ stands for the component of $\barPi$ belonging to $({\bf 3^*},{\bf 2^*},-1/6)$.  
The SU(3)-sextet and the SU(3)-singlet components of $\Pi$ are expressed by 
$\Pi_{(6)}$ and $\Pi_{(1)}$, while their conjugate representations in $\barPi$ are by 
$\barPi_{(6)}$ and $\barPi_{(1)}$.   
These superfields have masses of the order of $M_X$.  

     The mass terms for $P_j$ ($j$=1,2,3,4) give a non-vanishing mass to 
one linear combination of them.  
Taking $\{F_{ij}\}$ for a unitary matrix, the mass eigenstates of $P_j$ are 
expressed by $Q_i=F_{ij}P_j$.  
In these four linear combinations, one state $Q_4$ alone has a mass of the order 
of $M_X$, while the masses of the other three independent states are vanishing.  
Although some of the massive superfields further form mass terms with 
the massless superfields after SU(2)$\times$(1) breaking,  
the resultant mixings are negligible.   
We can regard $Q_i$ ($i$=1,2,3) as the superfields for SU(2)-doublet quarks. 
On the other hand, the superfields for SU(2)-singlet quarks are given by   
$(\overline{\bf 3},{\bf 1},-2/3)$ in $\Psi_i$ and $(\overline{\bf 3},{\bf 1},1/3)$ in $\barPsi_i$, 
which we express as $U^{ci}$ and $D^{ci}$, respectively.  
The superfields for SU(2)-doublet and SU(2)-singlet leptons are given by 
$({\bf 1},{\bf 2},-1/2)$ in $\barPsi_i$ and $({\bf 1},{\bf 1},1)$ 
in $\Psi_i$, being denoted respectively by $L^i$ and $E^{ci}$.  

     The superpotential for quark and lepton masses is given by 
\begin{eqnarray}
 W &=&  \eta_d^{ij} H_1\epsilon_2 Q_iD^c_j 
             + \eta_u^{ij} H_2\epsilon_2 Q_iU^c_j  
           \nonumber \\
             &+&   \eta_e^{ij} H_1\epsilon_2 L_iE^c_j 
    + \frac{1}{2}\kappa^{mij}(\epsilon_2 L_i) T_m\epsilon_2 L_j,  
\label{superpotentialM} 
\end{eqnarray}
where the terms containing heavy matter superfields have been neglected.  
The mass matrices for $d$-type quarks, $u$-type quarks, charged leptons, and 
neutrinos are given respectively by $M_d=\eta_dv_1/\sqrt{2}$, $M_u=-\eta_uv_2/\sqrt{2}$, 
$M_e=\eta_ev_1/\sqrt{2}$, and $M_\nu=\kappa^mv_{Tm}/\sqrt{2}$.  
At SU(5) breaking, the coefficients are expressed in terms of those for the superpotentials 
$W_{M1}$, $W_{M2}$, and $W_{M3}$ in Eq. (\ref{superpotential}) as 
\begin{eqnarray}
     \eta_d^{ij} &=& -\frac{1}{\sqrt{2}}
               \left(\sum_{k=1}^3F_{ik}^*{\overline\eta}^{kj}_{\Psi} 
	       + F_{i4}^*\eta^{j}_{\Pi\Psi}\right),  
\nonumber	       \\
     \eta_u^{ij} &=& 4\sum_{k=1}^3F_{ik}^*\eta^{kj}_{\Psi},  
\label{eta}\\
     \eta_e^{ij} &=& -\frac{1}{\sqrt{2}}{\overline\eta}^{ji}_{\Psi},  
\nonumber	       \\
     \kappa^{mij} &=& 2\kappa_{\Psi}^{kij}A_{km}.    
\nonumber
\end{eqnarray}
Experimentally, the quark and lepton masses, CKM matrix, and MNS matrix 
are, in principle, measurable at the electroweak energy scale.  
The values of $\eta_d^{ij}$, $\eta_u^{ij}$, $\eta_e^{ij}$, and 
$\kappa^{mij}$ at SU(5) breaking can then be specified, taking into account 
their energy dependencies described by renormalization group equations. 

     The relations for the masses between $d$-type quarks and charged leptons 
at SU(5) breaking are different from those of the minimal model, owing to mixing 
of $\Pi$ and $\Psi_i$ in $W_{M1}$ and coupling of $\Pi$ and $\barPsi_i$ in $W_{M2}$.    
Without loss of generality, the coefficient matrix ${\overline\eta}_{\Psi}$ is  
taken for ${\overline\eta}_{\Psi}=-\sqrt{2}\eta_e^D$, where $\eta_e^D$ is defined by 
$\eta_e^D={\rm diag}(\eta_e^1,\eta_e^2,\eta_e^3)$, with $\eta_e^i$ being 
the eigenvalues of $\eta_e$ which are determined phenomenologically.  
From Eq. (\ref{eta}) the coefficients $\eta_d^{ij}$ are expressed as 
\begin{equation}
\eta_d^{ij}=\sum_{k=1}^3F_{ik}^*\eta_e^{Dkj}- \frac{1}{\sqrt{2}}F_{i4}^*\eta_{\Pi\Psi}^j.  
\label{etade}
\end{equation}
If the new matter superfields $\Pi$ and $\barPi$ are not introduced,  
the coefficients satisfy the equalities $\eta_d^{ij}=\eta_e^{Dij}$.   
The same eigenvalues for $\eta_d$ and $\eta_e$ are predicted, which lead to 
apparent inconsistency with experimental measurements.  
However, these equalities do not hold any more.  
Flexibility for the values of $F_{ij}$ and $\eta_{\Pi\Psi}^i$ makes it possible for 
the eigenvalues of $\eta_d$ to be consistent with the phenomenological values.   

     Left-handed neutrinos have Majorana masses compatible with experiments, 
which is traced back to the couplings of $\barPsi_i$, $\barPsi_j$, and $T_m$ of the 
superpotential $W_{M3}$ in Eq. (\ref{superpotential}) and non-vanishing small VEVs 
for the neutral scalar components of $T_m$ in Eq. (\ref{vevT}).  
Furthermore, the MNS matrix is not correlated with the CKM matrix.  
The neutrino generation mixing is prescribed by $\kappa_\Psi^{mij}$ and independent 
of $\eta_\Psi^{ij}$, $\overline\eta_\Psi^{ij}$, and $\eta_{\Pi\Psi}^i$ which 
determine the masses and mixings of the quarks.  
Note that contribution of either $T_1$ or $T_2$ is sufficient for inducing 
the neutrino masses.  

     We turn to discussion on whether grand unification of 
SU(3), SU(2), and U(1) gauge symmetries is realized.  
Although new superfields are very heavy, some of them should have 
masses which are smaller than the supposed unification scale $M_X$.  
We assume that the decomposed components of $T_m$ and $\barT_m$ and those 
of $\Pi$ and $\barPi$ have masses respectively around $M_{Tm}$ and around $M_\Pi$.  
From the renormalization group equations at one-loop level, 
the fine structure constants $\alpha_3$, $\alpha_2$, 
and $\alpha_1$ for SU(3), SU(2), and U(1) are given by, at an energy scale $\mu$,  
\begin{eqnarray}
    \frac{1}{\alpha_i(\mu)} &=& \frac{1}{\alpha_i(M_W)}  
                   - \frac{1}{2\pi}R_i(\mu) \nonumber \\
   &-& \frac{7}{2\pi}\left[\sum_{m=1}^2 \log\frac{\mu}{M_{Tm}} + \log\frac{\mu}{M_\Pi}\right],  
\label{constants}
\end{eqnarray}
where $\alpha_1$ represents the normalized value for SU(5).  
The contributions of the superfields other than $T_m$, $\barT_m$, $\Pi$, 
and $\barPi$ are expressed as $R_i(\mu)$, which are the same as the 
minimal supersymmetric standard model.    

     The energy dependencies of $\alpha_i$ in Eq. (\ref{constants}) lead to 
grand unification of gauge symmetries.  
The amounts of contribution from a pair of $\bf 15$ and $\overline{\bf 15}$ are the same 
for three gauge coupling constants.   
The contributions of the minimal supersymmetric standard model 
make the amounts of $1/\alpha_i(M_W)-R_i(\mu)/2\pi$ convergent at $\mu=M_X$.  
Therefore, at the unification scale $M_X$ for the minimal model, 
the gauge coupling constants are unified also in the present model.  
 
     The proton could decay through dimension-five operators mediated by SU(3)-triplet 
components $H_C$ and $\barH_C$ of the Higgs superfields $H$ and $\barH$.  
The superpotential for their couplings with quark and lepton superpfields is given by 
\begin{eqnarray}
 W &=& \zeta_{UD}^{ij} \epsilon_3\barH_C U^c_i D^c_j  
             + \zeta_{QL}^{ij} \barH_C Q_i\epsilon_2 L_j 
           \nonumber \\
             &+&  \zeta_{UE}^{ij} H_C U^c_i E^c_j 
             + \zeta_{QQ}^{ij} \epsilon_3 H_C Q_i\epsilon_2 Q_j,  
\end{eqnarray}
where the coefficients at SU(5) breaking are expressed in terms of  
those in Eq. (\ref{superpotential}) as 
\begin{eqnarray}
     \zeta_{UD}^{ij} &=& -\frac{1}{\sqrt{2}}
               {\overline\eta}^{ij}_{\Psi},  
\nonumber	                         \\
     \zeta_{QL}^{ij} &=& -\frac{1}{\sqrt{2}}
               \left(\sum_{k=1}^3F_{ik}^*{\overline\eta}^{kj}_{\Psi} 
	       - F_{i4}^*\eta^{j}_{\Pi\Psi}\right),  
\label{zeta}	                         \\
     \zeta_{UE}^{ij} &=& -4\eta^{ij}_{\Psi},  
\nonumber		   \\ 
     \zeta_{QQ}^{ij} &=& -2\sum_{k,l=1}^3F_{ik}^*F_{jl}^*
                             \eta^{kl}_{\Psi}.  
\nonumber
\end{eqnarray}
The lifetime of the proton is determined by these couplings and 
the generation mixing structure of squarks and sleptons.  
The latter depends not only on the superpotential but also on 
supersymmetry-soft-breaking terms \cite{bajc}, which are prescribed by 
a theoretical framework above the GUT energy scale.   

     Taking into consideration the correlations between the coefficients in Eq. (\ref{zeta}) 
and those in Eq. (\ref{eta}), the model could be constrained by 
experimental lower bounds on the proton lifetime.  
In the minimal model which does not contain the superfields $\Pi$ and $\barPi$, 
the coefficient matrices $\zeta_{UD}$ and $\zeta_{QL}$ are equal to $\eta_d$ 
while $\zeta_{UE}$ and $\zeta_{QQ}$ are proportional to $\eta_u$.  
As a result, magnitudes of the couplings for quark and lepton superfields with 
$H_C$ and $\barH_C$ are specified by the experimental values for 
the masses and mixings of the quarks.  
However, these correlations are now more relaxed, 
owing to the contributions of $F_{ij}$ and $\eta_{\Pi\psi}^i$.  
The contributions of supersymmetry-soft-breaking terms also become more flexible.  
The constraints from the proton decay are weaker than the minimal model.   

     Finally we discuss leptogenesis.  It is seen from the superpotential 
in Eq. (\ref{superpotentialM}) that the leptonic decay of the SU(2)-triplet  
particle $T_m$, with its branching ratio being large, changes lepton number by two unit.  
Sufficient difference for amount between $T_m$ and its   
anti-particle in the early universe could lead to non-vanishing lepton number which 
is converted into baryon asymmetry of the present universe.  
This difference is yielded by the decays of the SU(2)-triplet 
particle $\Phi$ through the interactions with $T_n$ and $\barT_m$ 
in Eq. (\ref{superpotentialH}) which violate $CP$ invariance. 
We assume the kinetic condition $2m_{T1}<2m_{T2}<m_\Phi$, which is 
implied from the magnitudes of the VEVs in Eq. (\ref{vevT}) for neutrino masses.  

     For definiteness, we examine the process in which the scalar component 
of $\Phi$, denoted by $\phi_\Phi$, decays into the fermion components 
of $T_m$ and $\barT_n$, denoted by $\psi_{Tm}$ and $\psi_{\barT n}$.  
Supersymmetry guarantees that thus obtained results are also valid for the corresponding 
processes in which the other components of superfields participate.  
The mass eigenstate for the fermions with the mass $m_{Tm}$ is given by  
$\psi_{m} = \psi_{Tm}+(\psi_{\barT m})^c$ $(m=1,2)$, 
where the superscript $c$ stands for charge conjugation.  
The decays $\phi_\Phi \to \psi_1+\overline{\psi_2}$ and $\phi_\Phi \to \overline{\psi_1}+\psi_2$ 
are induced  both at tree level and at one-loop level through final state interactions.  
The latter is mediated by $t$-channel exchange of $\phi_\Phi$ after the production 
of $ \psi_m$ and $\overline{\psi_n}$, with $m$ and $n$ being any combination for $m,n=1,2$.  

     The difference of produced amount between $\psi_m$ and $\overline{\psi_m}$ is 
evaluated by a decay rate asymmetry, which is calculated as   
\begin{eqnarray}
      A &=& \frac{\Gamma (\phi_\Phi \to\psi_1+\overline{\psi_2})-
                       \Gamma (\phi_\Phi \to\overline{\psi_1}+\psi_2)}
                       {\Gamma (\phi_\Phi \to\psi_1+\overline{\psi_2})+
                       \Gamma (\phi_\Phi \to\overline{\psi_1}+\psi_2)}  
                        \\    
      &=&-\frac{1}{16\pi}{\rm Im}[(\lambda_{\phi 2}^{22})^2]
    \frac{|\lambda_{\phi 2}^{12}|^2-|\lambda_{\phi 2}^{21}|^2}
           {|\lambda_{\phi 2}^{12}|^2+|\lambda_{\phi 2}^{21}|^2}
                   \frac{r\sqrt{1-4r}}{1-r},  
                   \nonumber  
\end{eqnarray}
with $r=(m_{T2}/m_\Phi)^2$.  
The mass of $\psi_1$ has been neglected. 
If the relevant coupling constants are not suppressed much, the possible range of 
the asymmetry is given by $|A|< 10^{-2}$.  
Since the branching ratio of $\phi_\Phi$ decaying into $\psi_1\overline{\psi_2}$ or  
$\overline{\psi_1}\psi_2$ is large, sizable difference for amount is yielded 
between $\psi_m$ and $\overline{\psi_m}$.   
It should be noted that the sign of lepton number produced by the decay 
of $\psi_1$ (and $\overline{\psi_1}$) 
is opposite to that by the decay of $\psi_2$ (and $\overline{\psi_2}$).  
However, the leptonic branching ratio is different between $\psi_1$ and $\psi_2$, depending 
on the coefficients $\overline\lambda^m$ in Eq. (\ref{superpotentialH}) and $\kappa^m$ 
in Eq. (\ref{superpotentialM}).  
In addition, lepton asymmetry induced by the decay of the heavier particle $\psi_2$ 
would be diluted by lepton-number violating interactions of the lighter particle $\psi_1$ 
before the interactions go out of thermal equilibrium.  
Leptogenesis could well occur in the early universe.  
Lepton number may also be induced in the decay processes of $T_1$ 
through off-diagonal self-energy terms \cite{ma}.  
Since the origin of $CP$ violation for this mechanism is 
different from that for the above discussed one, both mechanisms could coexist.  

     In conclusion, the masses and mixings of quarks and leptons are described 
realistically within the framework of grand unified model based on SU(5) and supersymmetry.  
All the particles belong to group representations of up to rank two.  
The Higgs particles introduced for neutrino masses generate also non-vanishing lepton number 
in the early universe.

\end{document}